\documentclass[aps,prb, amsfonts, amssymb, amsmath, superscriptaddress, notitlepage,twocolumn,10pt,nobalancelastpage]{revtex4-2}
\usepackage{graphicx} 

\usepackage{hyperref}
\usepackage{bm}
\usepackage{textgreek}
\usepackage{soul}
\usepackage{accents}
\hypersetup{ 
    colorlinks=true, 
    linktoc=all,     
    linkcolor=blue,  
    citecolor =blue
} 
\usepackage{amsmath}
\newcommand{\bk}{\bm{k}}
\newcommand{\be}{\bm{e}}
\newcommand{\bn}{\bm{n}}

\newcommand{\bq}{\mathbf{q}}

\newcommand{\bK}{\mathbf{K}}
\newcommand{\bQ}{\mathbf{Q}}

\newcommand{\bM}{\mathbf{M}}

\newcommand{\bg}{\bm{\gamma}}

\newcommand{\kvec}{\vec{k}}
\newcommand{\qvec}{\vec{q}}

\newcommand{\nn}{\nonumber}

\begin{document}
\title{{The Effects of Inter-Valley Coupling of Dirac fermions near Four Dimensions}}

\author{S. Thiagarajan}
\affiliation{London Centre for Nanotechnology, University College London, Gordon St., London, WC1H 0AH, United Kingdom}
\author{F. Kr\"uger}
\affiliation{London Centre for Nanotechnology, University College London, Gordon St., London, WC1H 0AH, United Kingdom}
\affiliation{ISIS Facility, Rutherford Appleton Laboratory, Chilton, Didcot, Oxfordshire OX11 0QX, United Kingdom}

\begin{abstract}
We analyze the criticality of an Ising Gross-Neveu-Yukawa (GNY) theory of $N_\psi$ Dirac fermion valleys with intra- and inter-valley interactions, using a renormalization-group analysis in $4-\epsilon$ space-time dimensions
to one-loop order. The conventional GNY fixed point of decoupled valleys is unstable against inter-valley fluctuations which are naturally present if the symmetry breaking is driven by short-ranged interactions. 
At the new  fixed point, the critical exponents differ from the conventional GNY universality. Most importantly, Lorentz invariance is broken due to interference effects resulting from the relative rotation between 
valley coordinate frames. In the limit of large $N_\psi$ the critical fixed point with finite inter-valley coupling remains stable but Lorentz invariance is asymptotically restored. 
\end{abstract}

\maketitle

\section{Introduction}

Dirac fermions were first introduced in high-energy physics as fields used to construct Lorentz-invariant Lagrangians within the Standard Model~\cite{El-Batanouny+2020}. Massless Dirac fermions 
also arise in condensed matter systems, such as graphene \cite{Novoselov+2005}, topological Weyl and Dirac semimetals in three-dimensional solids \cite{Armitage+18}, and in Dirac quantum spin 
liquids \cite{Wen02,Kitaev06}, where fermionic spinons emerge through fractionalization of spin degrees of freedom.  The emergence of Dirac fermions in the low-energy spectra of quantum material is a result 
of the underlying lattice symmetries, such as sublattice and point-group symmetries \cite{El-Batanouny+2020}.

In the presence of sufficiently strong short-ranged repulsive interactions, Dirac semimetals are known to become unstable towards symmetry breaking, resulting in the opening of a mass gap at the Dirac points.
The type of order depends on the nature of the microscopic interactions, e.g. for the half-filled Hubbard model on the honeycomb lattice with competing interactions, a vast array of phases were found \cite{Grushin+13}, 
including antiferromagnetism, different types of charge order, Kekule phases, and topological quantum Hall states. 

The critical behavior of such quantum phase transitions can be described by Gross-Neveu-Yukawa (GNY) field theories similar to that which describes chiral symmetry breaking and spontaneous mass generation 
in high-energy physics \cite{Gross+1974,Justin+1991}. In the condensed matter setting \cite{Herbut+2006,Herbut+2009,Herbut+2014}, the bosonic order parameter field is introduced through a Hubbard-Stratonovich 
decoupling of a four-fermion vertex in the appropriate channel. The resulting GNY theories can differ by the dimensionality,  type of Yukawa coupling in the fermion-flavor space and by the number of order-parameter 
components. 

The Nielsen-Ninomiya theorem requires chiral fermions on a lattice to appear in pairs~\cite{Nielsen+1981,Suzuki+2004}. However, this theorem relies on assumptions including locality, hermiticity, periodicity, 
bilinearity, and chiral symmetry~\cite{Nielsen+1981}. Allowing long-range hopping relaxes the locality assumption. For example, SLAC fermions \cite{Drell+1976,Lang+19} provide a construction in 
which the fermion-doubling problem is circumvented while chiral symmetry remains preserved. 

\begin{figure}[t!]
    \centering
    \includegraphics[width=\linewidth]{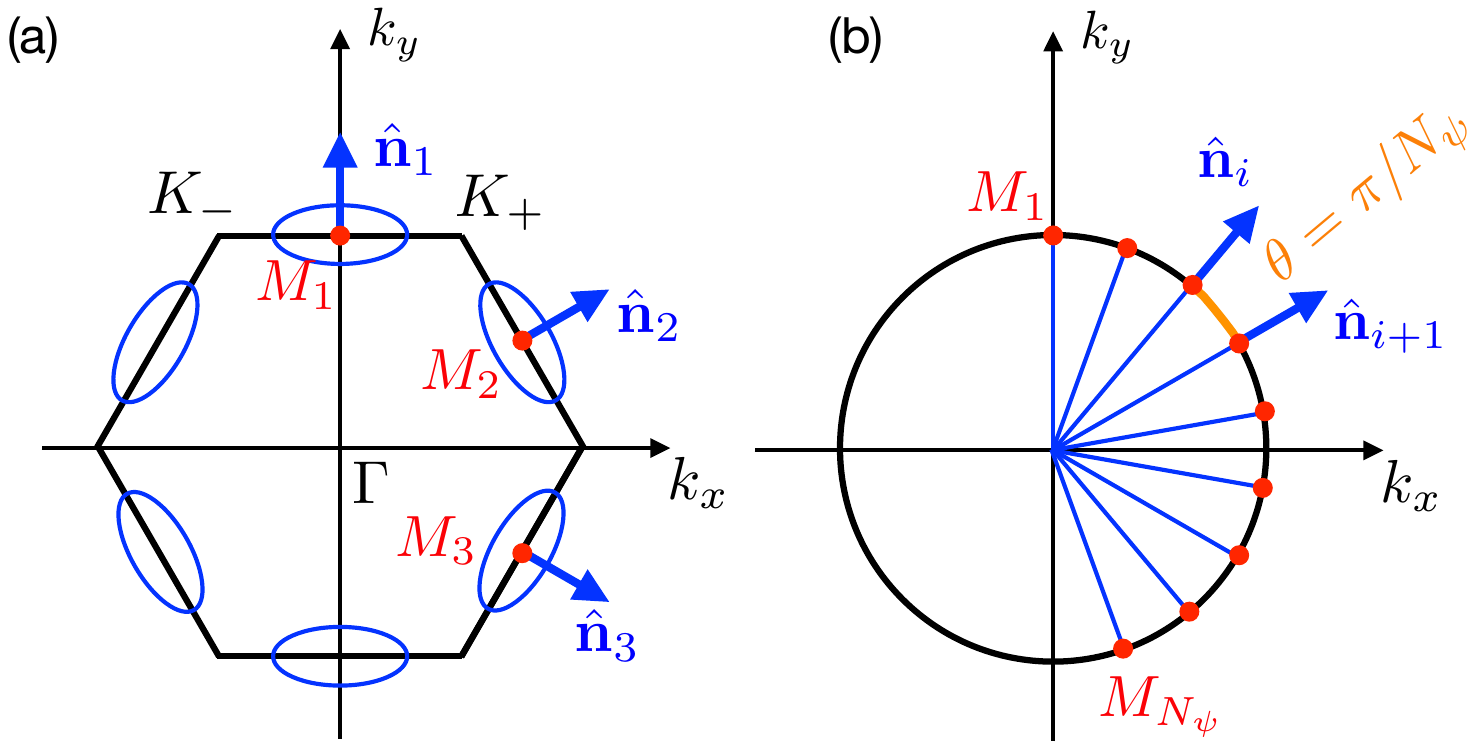}
    \caption{(a) Hexagonal Brillouin zone with Dirac valleys at the three $M$ points, as observed at the topological phase transition of the Kitaev QSL  in magnetic field along the $[111]$ 
    direction \cite{Ralko+20,Yilmaz+22,Zhang+2022,Thiagarajan+2026}. The directors $\hat{\bn}_i$ determine the orientation of the local coordinate frames. (b) Generalization to $N_\psi$ Dirac points, where the directors 
    of neighboring valleys enclose an angle of $\theta = \pi/N_\psi$.}
    \label{figure1}
\end{figure}

Furthermore, recent studies \cite{Ralko+20,Yilmaz+22,Zhang+2022,Thiagarajan+2026} of the Kitaev honeycomb model in a magnetic field along the $[111]$ direction showed that at the topological phase transition 
the spinon gap closes at the three $M$ points of the hexagonal Brillouin zone [see Fig.~\ref{figure1}(a)]. Using a strong-coupling renormalization-group (RG) analysis of the critical GNY field theory with 
three Dirac valleys and both intra- and inter-valley fluctuation fields \cite{Thiagarajan+2026}, it was shown that the inter-valley interactions change the universality of the transition. Most importantly, they lead 
to a breaking of Lorentz invariance through interference effects caused by the different relative orientations of the local coordinate frames.

In the study of symmetry-breaking phase transitions of Dirac semimetals the coupling between valleys is usually neglected and the valley index is simply included in  the number of featureless fermion flavors, 
justifying large $N$ approximations. However, the quantum phase transitions  are typically driven by short-range interactions such as local Hubbard repulsions. The locality of interactions on the lattice will 
necessarily lead to inter-valley coupling in momentum space.  

In this article we will investigate an Ising GNY field theory of $N_\psi$ Dirac valleys with both intra- and inter-valley fluctuation fields. We use an $\epsilon$ expansion below $3+1$ space-time dimensions to systematically 
study the RG flow in the space of the two Yukawa couplings and relevant bosonic interaction vertices. Our results show that the Lorentz invariant GNY fixed point with decoupled valleys is unstable against 
inter-valley coupling, resulting in a new critical fixed point where both Yukawa couplings are finite and Lorentz invariance is broken.      

The remainder of this paper is organized as follows. In Sec.~\ref{sec.FT} we introduce the Ising GNY theory of $N_\psi$ Dirac valleys with bosonic inter- and intra-valley fluctuation fields and various bosonic interaction 
vertices. In Sec.~\ref{sec.RG} we present our RG analysis of the theory. We derive expressions for the anomalous dimensions of the fields and the dynamical critical exponent,  and 
RG equations for the boson velocities, Yukawa couplings and bosonic vertices in $4-\epsilon$ dimensions to one-loop order. Finally, we derive the RG equations for the mass terms of the fluctuation fields, allowing us to 
compute correlation-length exponents.  The results deduced from the RG analysis are presented in Sec.~\ref{sec.results} and discussed in Sec.~\ref{sec.discussion}.

\section{Effective Field Theory}
\label{sec.FT}

Our model is motivated by the low-energy behavior at the field-driven topological phase transition of the Kitaev model, which shows three Dirac valleys at the $M$ points in the hexagonal Brillouin zone, as illustrated 
in Fig.~\ref{figure1}(a), and necessarily contains both intra- and inter-valley interactions \cite{Thiagarajan+2026}. Here we will study a continuum field theory with an arbitrary number of $N_\psi\ge 3$ coupled valleys
[Fig.~\ref{figure1}(b)]. The orientations of the local coordinate frames of the valleys are described by the  directors $\hat{\bn}_i$. We introduce an additional relativistic momentum direction to facilitate a $4-\epsilon$ 
expansion. 

The zero-temperature free Dirac fermion action is given by 
\begin{eqnarray}
S_0[\bar{\bm\psi},\bm\psi] & = &  \sum_{i=1}^{N_\psi}\int_{\vec{k}} \bar{\bm\psi}_i(\vec{k})  \Big\{-i k_0  +  [(\hat{\bn}_i \times  \hat{\be}_3) \cdot  \bk] \bg_1\nn\\
& & +  (\hat{\bn}_i \cdot \bk) \bg_2 + (\hat{\be}_3\cdot\bk) \bg_3 \Big\}\bm\psi_i(\vec{k}),
\end{eqnarray}
 where $\vec{k}=(k_0,\bk)^T$ is the four-component frequency-momentum vector,   $i$ labels the Dirac valleys, and we use gamma matrices $\bg_\mu$ in pseudo-spin space. 
The valleys  have rotated coordinate frames to one another, described by directors $\hat{\bn}_i = (-\sin \theta_i,\cos\theta_i,0)^T$ 
 in the $k_x$-$k_y$ plane, where $\theta_i=i \pi/N_\psi$, and $\hat{\be}_3 = (0,0,1)^T$ is the unit vector along $k_z$.
 
We will next consider a Yukawa coupling of the Dirac fermions to an Ising order parameter field.  Since the interactions are short ranged, both inter- and intra-valley fluctuations will be important. On the full two-dimensional Brillouin 
zone the fluctuations in the mass channel are of the form $\Phi(\bQ)\bar{\bm{\psi}}(\bK)\bg_4 \bm{\psi}(\bK+\bQ)$, where we have dropped the dependence on frequencies for brevity.  In the low-energy theory, we consider momentum 
patches close to the $M$ points, 
$\bK=\bM_i+\bk$, and define  $\bm{\psi}(\bK) = \bm{\psi}(\bM_i+\bk)\equiv \bm{\psi}_i(\bk)$. 
For the  momentum transfer we write $\bQ=\bQ_{ij}+\bq$ where $\bQ_{ij} = \bM_j-\bM_i$ and $\bq$ is small. Defining the intra-valley fluctuation fields ($i=j$) as $\Phi(\bQ_{ii}+\bq)=\Phi(\bq)=\phi(\bq)$ and the 
inter-valley fluctuation fields ($i\neq j$) as $\Phi(\bQ_{ij}+\bq)=\varphi_{ij}(\bq)$, the Yukawa couplings can be written as
 \begin{eqnarray}
 S_g & = &  \frac{g_\phi}{\sqrt{N_\psi}} \sum_{i=1}^{N_\psi} \int_{\vec{k},\vec{q}} \phi(\vec{q}) \, \bar{\bm\psi}_i(\vec{k}) \bg_4 \bm\psi_i(\vec{k}+ \vec{q}),\\
 & & +  g_\varphi \sum_{i=1}^{N_\psi} \int_{\vec{k},\vec{q}}   \Big\{ \varphi_{i,i+1}(\vec{q}) \,  \bar{\bm\psi}_i(\vec{k}) \bg_4 \bm\psi_{i+1}(\vec{k}+ \vec{q})\nn\\
& &   \quad+  \varphi_{i+1,i}(\vec{q}) \,  \bar{\bm\psi}_{i+1}(\vec{k}) \bg_4 \bm\psi_{i}(\vec{k}+ \vec{q}) \Big\}.
 \end{eqnarray}
We have rescaled the intra-valley Yukawa coupling $g_\phi$ by a factor of $\sqrt{N_\psi}$ 
since the same boson field couples to all valleys. We further assume that only neighboring  
 valleys are coupled to reduce the number of coupling constants. The bosonic fluctuation fields satisfy  $\phi^*(\kvec)=\phi(-\kvec)$ and $\varphi_{i,i+1}^*(\kvec)=\varphi_{i+1,i}(-\kvec)$.
 The quadratic actions for the bosonic fluctuation fields are given by 
\begin{eqnarray}
S_0[\phi,\varphi] & = & \frac{1}{2}\int_{\vec{q}} \left[q_0^2+c_\phi^2(q_1^2+q_2^2)+q_3^2\right]  |\phi(\vec{q})|^2\\
&  & + \frac{1}{2} \sum_{i=1}^{N_\psi} \int_{\vec{q}} \left[q_0^2+c_\varphi^2(q_1^2+q_2^2)+q_3^2\right]  |\varphi_{i,i+1}(\vec{q})|^2.\nn
\end{eqnarray} 
The independent in-plane velocities $c_\phi$ and $c_\varphi$ are required to account for the interference effects due to the rotation of local coordinate frames. The quartic interaction vertices of the 
 bosonic fluctuation fields are given by
 \begin{eqnarray}
  S_\lambda & = & \int_{\vec{k}_1,\vec{k}_2,\vec{k}_3,\vec{k}_4} \delta( \vec{k}_1+\vec{k}_2+\vec{k}_3+\vec{k}_4)  \nn\\
  & & \times\Big\{ \lambda_\phi \,\phi(\vec{k}_1)  \phi (\vec{k}_2)\phi(\vec{k}_3)  \phi(\vec{k}_4)\nn\\
  & &  + \lambda_{\varphi}  \sum_{i=1}^{N_\psi}  \varphi_{i,i+1}(\vec{k}_1) \varphi_{i+1,i}(\vec{k}_2) \varphi_{i,i+1}(\vec{k}_3) \varphi_{i+1,i}(\vec{k}_4) \nn\\
  & &  +\tilde{\lambda} \sum_{i=1}^{N_\psi}  \phi(\vec{k}_1) \phi(\vec{k}_2)\varphi_{i,i+1}(\vec{k}_3) \varphi_{i+1,i}(\vec{k}_4)\Big\}.
 \end{eqnarray}

\section{Renormalization Group  Analysis}
\label{sec.RG}

We perform a renormalization-group (RG) analysis of the action $S[\bar{\bm\psi},\bm\psi,\phi,\varphi] = S_0[\bar{\bm\psi},\bm\psi] + S_0[\phi,\varphi] +  S_g[\bar{\bm\psi},\bm\psi,\phi,\varphi] + S_\lambda[\phi,\varphi]$
to one-loop order to obtain the scale dependence of the Yukawa couplings $g_\phi$ and $g_\varphi$, the velocities $c_\phi$ and $c_\varphi$, and the bosonic
interaction vertices $\lambda_\phi$, $\lambda_\varphi$ and $\tilde{\lambda}$. To do so, we integrate out UV modes from the infinitesimal frequency-momentum shell below the cut-off $\Lambda$, 
\begin{equation}
    \Lambda e^{-d\ell} \leq |\vec{q}| \leq \Lambda,
    \label{eq.shell}
\end{equation}
 followed by a rescaling of frequency and momenta,
\begin{equation}
k_{1,2} \rightarrow e^{-d\ell} k_{1,2}, \; k_{0,3} \rightarrow e^{-z d\ell} k_{0,3},
\end{equation}
and fields, 
\begin{eqnarray}
\bm\psi \rightarrow e^{-\Delta_\psi/2d\ell} \bm\psi , \; \phi \rightarrow e^{-\Delta_\phi/2d\ell} \phi, \; \varphi \rightarrow e^{-\Delta_\varphi/2d\ell} \varphi.
\end{eqnarray}

The momentum component $k_3$, perpendicular to the plane of Dirac valleys, scales in the same way as frequency $k_0$, while $z$ denotes the dynamical 
exponent. 

\begin{figure}[t!]
    \centering
    \includegraphics[width=\linewidth]{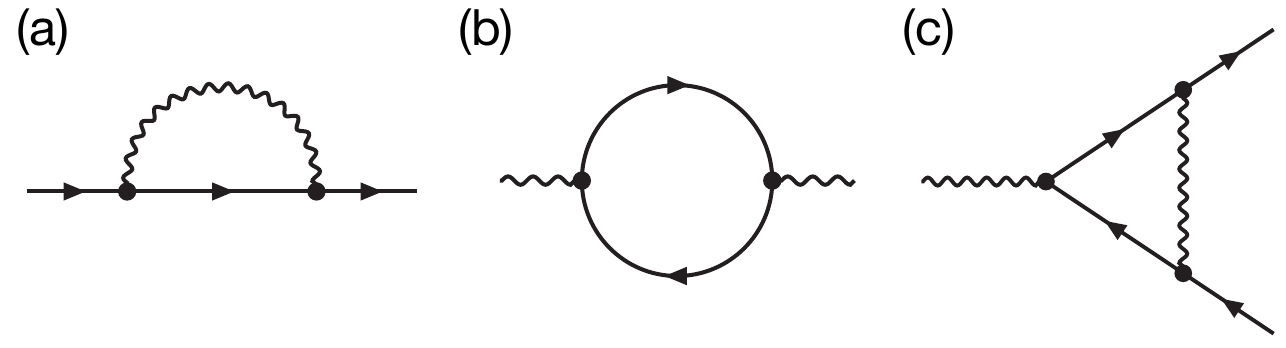}
    \caption{Feynman diagrams that correspond to (a) fermionic self-energy corrections, (b) bosonic self-energy corrections, and (c) the renormalization of the Yukawa vertices. The solid lines represent 
    fermionic Grassmann fields, wiggly lines bosonic intra- and inter-valley fluctuation fields.}
    \label{figure2}
\end{figure}

 \subsection{Fermion Self-Energy Corrections}
 
 We start by computing the fermion self-energy corrections $d\bm{\Sigma}_i$ to the inverse fermion propagator at valley $i$, corresponding to the 
 diagram shown in Fig.~\ref{figure2}(a). The contributions from intra- and inter-valley fluctuations are given by the following integrals over the infinitesimal 
 frequency-momentum shell defined in Eq. (\ref{eq.shell}), 
 \begin{eqnarray}
 d\bm{\Sigma}_i (\kvec) & = &  -\frac{{g_{\phi}}^2}{N_\psi} \int_{\qvec}^{>} G_\phi(\qvec) \bg_4 \bm{G}_i (\kvec + \qvec) \bg_4  \\
 & & -g_\varphi^2 \int_{\qvec}^{>} G_\varphi(\qvec) \bg_4 [\bm{G}_{i-1}+\bm{G}_{i+1}](\kvec + \qvec) \bg_4, \nn
\end{eqnarray}
where 
\begin{equation}
\bm{G}_i (\kvec) = \frac{i k_0+[(\hat{\bn}_i \times  \hat{\be}_3) \cdot  \bk] \bg_1+  (\hat{\bn}_i \cdot \bk) \bg_2 + (\hat{\be}_3\cdot\bk) \bg_3 }{\kvec^2}
\end{equation}
is the fermion Green function in valley $i$ and 
\begin{equation}
G_{\phi/\varphi} (\qvec) = \frac{1}{q_0^2+c_{\phi/\varphi}^2(q_1^2+q_2^2)+q_3^2}
\end{equation}
are the bosonic fluctuation propagators. Performing a Taylor expansion in outer frequency and momenta $\kvec$ to linear order, the fermion self-energy can be written 
in the form
\begin{eqnarray}
 d\bm{\Sigma}_i (\kvec) & = &  -i k_0 \Sigma_0 d\ell + [(\hat{\bn}_i \times  \hat{\be}_3) \cdot  \bk] \bg_1 \Sigma_1 d\ell\nn\\
 & & +  (\hat{\bn}_i \cdot \bk) \bg_2 \Sigma_2 d\ell + (\hat{\be}_3\cdot\bk) \bg_3 \Sigma_3 d\ell,
\end{eqnarray}
with shell integrals $\Sigma_\mu d\ell$ that can be easily computed using four-dimensional spherical coordinates, $q_0=q \cos\phi_1$, $q_3=q \sin\phi_1\cos\phi_2$, 
$q_1 = q\sin\phi_1\sin\phi_2\cos\vartheta$, and $q_2 = q\sin\phi_1\sin\phi_2\sin\vartheta$ over the range $\phi_1,\phi_2\in[0,\pi]$, $\vartheta\in[0,2\pi]$ and $q\in[\Lambda e^{-d\ell},\Lambda]$
with volume element $dV=q^3 \sin^2\phi_1\sin\phi_2 \, d\phi_1\, d\phi_2\, d\vartheta \, dq$.
The resulting self-energy coefficients are 
\begin{eqnarray}
\Sigma_0 & = & \Sigma_3 = \frac{{g_\phi}^2\Lambda_c}{N_\psi} f_{0}(c_\phi^2) + 2 g_\varphi^2 \Lambda_c f_{0}(c_\varphi^2), \\
\Sigma_1 & = & \Sigma_2 = \frac{{g_\phi}^2\Lambda_c}{N_\psi} f_{1}(c_\phi^2) + 2 g_\varphi^2 \Lambda_c f_{1}(c_\varphi^2) \cos(\theta).
\end{eqnarray}
The inter-valley contribution to $\Sigma_{1,2}$ contains a factor $\cos(\theta)$, where $\theta$ is the angle between neighboring Dirac valleys. This 
arises from the relation $\hat{\bn}_{i-1}+\hat{\bn}_{i+1} = 2 \cos(\theta)\hat{\bn}_i$. 

For brevity, we have defined $\Lambda_c = S_4/(2\pi)^4$ with $S_4=2\pi^2$ the surface area of the four-dimensional 
unit sphere. This factor appears in all shell integrals and can be scaled out of the theory by rescaling the Yukawa couplings, $g^2 \to g^2/ \Lambda_c$
and bosonic interactions, $\lambda\to\lambda/  \Lambda_c$, which we will use in the following.

The functions $f_0$ and $f_1$ are defined through two-dimensional angular integrals, e.g. 
\begin{eqnarray}
f_0(c^2) & = &  \frac{1}{\pi} \int_0^\pi d\phi_1 \int_0^\pi d\phi_2 \frac{\sin^4\phi_1 \sin^3\phi_2}{1-(1-c^2)\sin^2\phi_1 \sin^2\phi_2}\nn\\ 
& & = \frac{-(1-c^2)-\ln c^2}{(1-c^2)^2},
\end{eqnarray}
and $f_1(c^2) = 1 - c^2 f_0(c^2)$. Note that $\lim_{c^2\to1} f_i (c^2) = 1/2$.

Under RG our free fermion action must remain invariant after including the loop corrections and the rescaling of frequency, momenta and fields. This leads to the simultaneous equations
\begin{eqnarray}
  -3z - 2 - \Delta_\psi + \Sigma_0 & = & 0,\\
  -2z - 3 - \Delta_\psi  + \Sigma_1 & = & 0,
\end{eqnarray}
from which we obtain the dynamical exponent, 
\begin{eqnarray}
z & = & 1 +  \Sigma_0 - \Sigma_1 \nn\\
& = & 1 + \frac{{g_\phi}^2}{N_\psi} \left[f_{0} (c^2_\phi) - f_{1} (c^2_\phi) \right]\nn\\
& & + 2g_\varphi^2 \left[ f_{0}(c^2_\varphi) -f_{1}(c^2_\varphi)\cos(\theta)\right],
\label{eq.z}
\end{eqnarray}
and the fermion anomalous dimension, 
\begin{eqnarray}
\eta_\psi & = & 3 \Sigma_1 - 2 \Sigma_0 \nn\\
& = & \frac{{g_\phi}^2}{N_\psi} \left[3f_{1} (c^2_\phi) - 2f_{0} (c^2_\phi) \right]\nn\\
& & + 2 g_\varphi^2 \left[ 3f_{1}(c^2_\varphi) \cos(\theta)  - 2f_{0}(c^2_\varphi)\right],
\label{eq.etapsi}
\end{eqnarray}
where the fermion field rescales with $\Delta_\psi = -5+\eta_\psi$.

\subsection{Boson Propagator Renormalization}

The corrections to the boson propagators are described by the Feynman Diagrams in Fig.~\ref{figure2}(b). The resulting self-energy corrections to the propagators of the intra-
and inter-valley fluctuations are given by
\begin{eqnarray}
\Pi_\phi(\qvec)d\ell & = & \frac{g_\phi^2}{N_\psi  \Lambda_c}\sum_{i=1}^{N_\psi} \int_{\kvec}^> \bm{G}_i (\kvec) \bg_4 \bm{G}_i (\kvec + \qvec) \bg_4 \\
& = &  2 g_\phi^2  |\qvec|^2 d\ell,\nn\\
\Pi_\varphi(\qvec)d\ell & = & 2 \frac{g_\varphi^2}{ \Lambda_c}  \int_{\kvec}^> \bm{G}_i (\kvec) \bg_4 \bm{G}_{i+1} (\kvec + \qvec) \bg_4 \\
& = & 4g_{\varphi}^2\left[ q_0^2+  \cos(\theta) (q_1^2+q_2^2) + q_3^2\right] d\ell, \nn
\end{eqnarray}
where we have performed a Taylor expansion to second order in the external frequency and momenta $\qvec$ and subtracted the $\qvec=0$ contribution 
for regularization. Contractions of the bosonic interaction vertices only contribute to the mass renormalization of the fluctuation fields, which will be considered later.

Combining with the rescaling and requiring that the coefficients of the $q_0^2$ and $q_3^2$ terms in the propagators remain constant under the RG, we obtain 
$\Delta_\phi = -4z -2 + 2 g_\phi^2$ and $\Delta_\varphi = -4z -2 + 4 g_\varphi^2$ for the scaling dimensions of the intra- and inter-valley fluctuation fields. 
From Eq.~(\ref{eq.z}) for the dynamical exponent $z$ it follows that $\Delta_{\phi/\varphi} = -6 +\eta_{\phi/\varphi}$ with anomalous dimensions
\begin{eqnarray}
\label{eq.etaphi}
\eta_\phi & = &  2 g_\phi^2 -4 \frac{{g_\phi}^2}{N_\psi} \left[f_{0} (c^2_\phi) - f_{1} (c^2_\phi) \right]\\
& & -8 g_\varphi^2 \left[ f_{0}(c^2_\varphi) -f_{1}(c^2_\varphi)\cos(\theta)\right],\nn\\
\label{eq.etavarphi}
\eta_\varphi & = & 4 g_\varphi^2 -4 \frac{{g_\phi}^2 }{N_\psi} \left[f_{0} (c^2_\phi) - f_{1} (c^2_\phi) \right]\\
& & -8 g_\varphi^2 \left[ f_{0}(c^2_\varphi) -f_{1}(c^2_\varphi)\cos(\theta)\right].\nn\
\end{eqnarray}

With the above conventions, the velocity renormalization is determined by the RG equations
\begin{eqnarray}
\label{eq.cphi}
\frac{d c_\phi^2}{d\ell} & = & 2(z-1) c_\phi^2+2 g_\phi^2 (1-c_\phi^2),\\
\frac{d c_\varphi^2}{d\ell} & = & 2(z-1) c_\varphi^2+4 g_\varphi^2  \left[\cos(\theta)-c_\varphi^2\right],
\label{eq.cvarphi}
\end{eqnarray}
where $z=z(c_\phi^2,c_\varphi^2,g_\phi,g_\varphi)$ is defined in Eq.~(\ref{eq.z}).

\begin{figure}[t!]
    \centering
    \includegraphics[width=\linewidth]{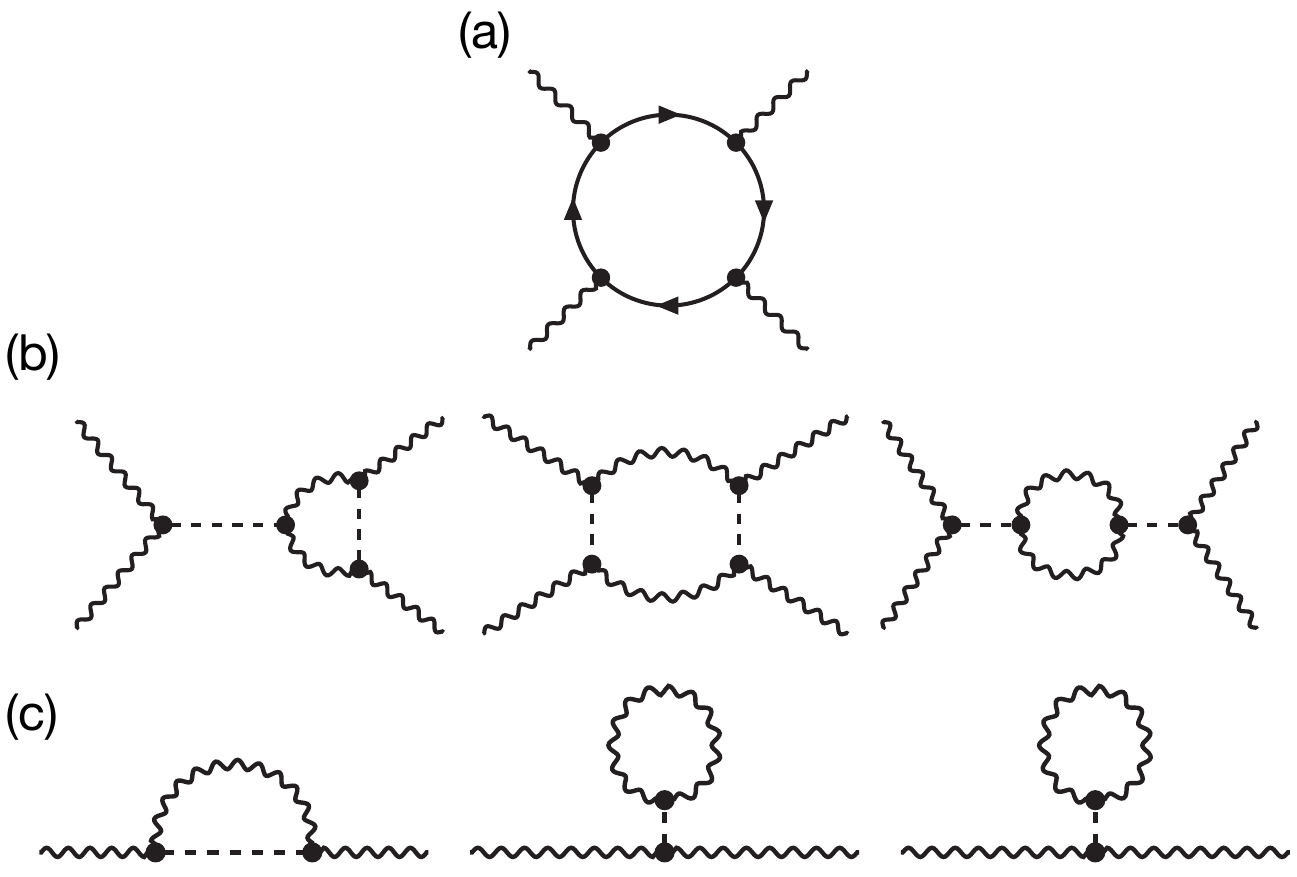}
    \caption{Feynman diagrams that correspond to the renormalization of bosonic interaction vertices by the (a) Yukawa couplings and (b) the bosonic interactions.  
    (c) Diagrams that generate mass terms of the bosonic fluctuation fields.}
    \label{figure3}
\end{figure}

\subsection{Renormalization of the Yukawa Couplings}

The Yukawa couplings are renormalized by diagrams shown in Fig.~\ref{figure2}(c). The loop correction to the intra-valley Yukawa coupling $g_\phi$ is given by
\begin{eqnarray} 
d g_\phi & = & \frac{g_\phi^3}{N_\psi \Lambda_c}  \int_{\qvec}^> G_\phi(\qvec) \bm{G}_i (\qvec) \bg_4 \bm{G}_i (\qvec) \bg_4\\
& & + \frac{2 g_\phi g_\varphi^2}{3 \Lambda_c}  \int_{\qvec}^> G_\varphi(\qvec) \bm{G}_i (\qvec) \bg_4 \bm{G}_i (\qvec) \bg_4\nn\\
& = & -\frac{g_\phi^3}{N_\psi} \left[f_0(c_\phi^2)+f_1(c_\phi^2)\right]d\ell \nn\\
& &  -\frac{2 g_\phi g_\varphi^2}{3}  \left[f_0(c_\varphi^2)+f_1(c_\varphi^2)\right]d\ell,\nn
\end{eqnarray}
while the corrections for the  inter-valley Yukawa coupling $g_\varphi$ is
\begin{eqnarray} 
d g_\varphi & = & \frac{g_\varphi^3}{2\Lambda_c}  \int_{\qvec}^> G_\varphi(\qvec) \bm{G}_i (\qvec) \bg_4 (\bm{G}_{i-1} +\bm{G}_{i+1} )(\qvec) \bg_4\\
& & + \frac{g_\phi^2 g_\varphi}{6 N_\psi\Lambda_c}  \int_{\qvec}^> G_\phi(\qvec) \bm{G}_i (\qvec) \bg_4 (\bm{G}_{i-1} +\bm{G}_{i+1} ) (\qvec) \bg_4\nn\\
& = & -g_\varphi^3 \left[f_0(c_\varphi^2)\cos(\theta)+f_1(c_\varphi^2)\right]d\ell\nn\\
& & - \frac{g_\phi^2 g_\varphi}{3 N_\psi} \left[f_0(c_\phi^2)\cos(\theta)+f_1(c_\phi^2)\right]d\ell.\nn
\end{eqnarray}

Combining with the rescaling and inserting the expressions for $z$ (\ref{eq.z}), $\eta_\psi$ (\ref{eq.etapsi}), $\eta_\phi$ (\ref{eq.etaphi}) and $\eta_\varphi$ (\ref{eq.etavarphi}), we obtain the following 
RG equations in $4-\epsilon$ space-time dimensions, 
\begin{eqnarray}
\label{eq.gphi}
\frac{d g_\phi^2}{d\ell} & = & \epsilon g_\phi^2 - 2 \left[1+\frac{f_0(c_\phi^2)+2 f_1(c^2_\phi)}{N_\psi}\right]g_\phi^4\\
& & - \frac{4}{3}\left[    f_0(c_\varphi^2) + f_1(c_\varphi^2)    \left[ 1+3\cos(\theta)\right]    \right]  g_\phi^2 g_\varphi^2, \nn\\
\label{eq.gvarphi}
\frac{d g_\varphi^2}{d\ell} & = & \epsilon g_\varphi^2  -\frac{2}{3 N_\psi} \left[ f_0(c_\phi^2) \cos(\theta) +4 f_1(c_\phi^2)    \right] g_\phi^2 g_\varphi^2\\
& & -2 \left[ 2+ f_0(c_\varphi^2)\cos(\theta) + f_1(c_\varphi^2) \left[ 1+2\cos(\theta)\right]   \right] g_\varphi^4.\nn
\end{eqnarray}

Note that the bosonic interaction vertices $\lambda_\phi$, $\lambda_\varphi$, and $\tilde\lambda$ don't contribute to the renormalization of the Yukawa couplings and critical 
boson propagators.

\subsection{Renormalization of Boson Vertices}

The renormalization of the quartic bosonic interaction vertices is given by the diagrams shown in Fig~\ref{figure3} (a) and (b). Their calculation is straightforward and results in the following set 
of coupled RG equations, 
\begin{eqnarray}
\label{eq.lambda1}
\frac{d\lambda_\phi}{d\ell} & = & \epsilon \lambda_\phi+2(z-1)\lambda_\phi-4 g_\phi^2\lambda_\phi\\
& & +\frac{1}{N_\psi}g_\phi^4-\frac{36}{c_\phi^2} \lambda_\phi^2-\frac{N_\psi}{2 c_\varphi^2}\tilde{\lambda}^2,\nn\\
\label{eq.lambda2}
\frac{d\lambda_\varphi}{d\ell} & = & \epsilon \lambda_\varphi+2(z-1)\lambda_\varphi-8 g_\varphi^2\lambda_\varphi\\
& & + \frac13 \left[\cos(2\theta)+4\cos(\theta)- 1 \right] g_\varphi^4-\frac{10}{c_\varphi^2}\lambda_\varphi^2 -\frac{1}{c_\phi^2}\tilde{\lambda}^2,\nn\\
\label{eq.lambda3}
\frac{d\tilde{\lambda}}{d\ell} & = & \epsilon \tilde{\lambda} +2(z-1)\tilde{\lambda}-2 (g_\phi^2+2g_\varphi^2)\tilde{\lambda}\\
& & + \frac{2}{3N_\psi}\left[ \cos(\theta)+3  \right]g_\phi^2 g_\varphi^2 -\frac{6}{c_\phi^2}\lambda_\phi\tilde{\lambda}-\frac{2}{c_\varphi^2}\lambda_\varphi\tilde{\lambda}\nn\\
& & -4 \rho(c_\phi^2,c_\varphi^2) \tilde{\lambda}^2,\nn
\end{eqnarray}
where $z=z(c_\phi^2,c_\varphi^2,g_\phi,g_\varphi)$, as defined in Eq.~(\ref{eq.z}). The function 
\begin{equation}
\rho(c_\phi^2,c_\varphi^2) = \frac{\ln(c_\phi^2/c_\varphi^2)}{c_\phi^2-c_\varphi^2}
\end{equation}
in the RG equation for $\tilde{\lambda}$ arrises from angular integration. Note that in the limit $c_\varphi^2 \to c_\phi^2$ the function approaches the value $1/c_\phi^2$.

\subsection{Mass Renormalization and Correlation-Length Exponents}

Finally, we compute the renormalization of the masses $m_\phi^2$ and $m_\varphi^2$ of the fluctuation fields by the bosonic interaction vertices, corresponding to the diagrams
shown in Fig.~\ref{figure3}(c). Combining the loop corrections with the rescaling, to leading order, we obtain coupled RG equations
\begin{equation}
\label{eq.mass}
\frac{d}{d\ell}\left(\begin{array}{c}  m_\phi^2 \\ m_\varphi^2 \end{array}\right) =  \left(\begin{array}{cc} 2 - \eta_\phi -  \frac{12\lambda_\phi}{c_\phi^2} & -\frac{\tilde{\lambda}}{c_\varphi^2}\\  -\frac{\tilde{\lambda}}{c_\phi^2} & 2- \eta_\varphi  - \frac{8\lambda_\varphi}{c_\varphi^2}\end{array}\right) \left(\begin{array}{c}  m_\phi^2 \\ m_\varphi^2 \end{array}\right),
\end{equation}
where the anomalous dimensions $\eta_\phi$ and $\eta_\varphi$ are given in Eqs. (\ref{eq.etaphi}) and (\ref{eq.etavarphi}), respectively.  The eigenvalues of the above 2x2 coefficient matrix, evaluated at the critical fixed point, 
correspond to the inverse correlation-length exponents $1/\nu_1$ and $1/\nu_2$.

\section{Results}
\label{sec.results}

Since the bosonic interaction vertices $\lambda_\phi$, $\lambda_\varphi$, and $\tilde\lambda$ do not contribute to the renormalization of the Yukawa couplings or critical 
boson propagators, we first analyze the RG flow and fixed point behavior of the RG equations for the boson velocities 
$c_\phi^2$ (\ref{eq.cphi}), $c_\varphi^2$ (\ref{eq.cvarphi}) and Yukawa couplings $g_\phi^2$ (\ref{eq.gphi}), $g_\varphi^2$ (\ref{eq.gvarphi}). 

\begin{figure}[t!]
    \centering
    \includegraphics[width=0.9\linewidth]{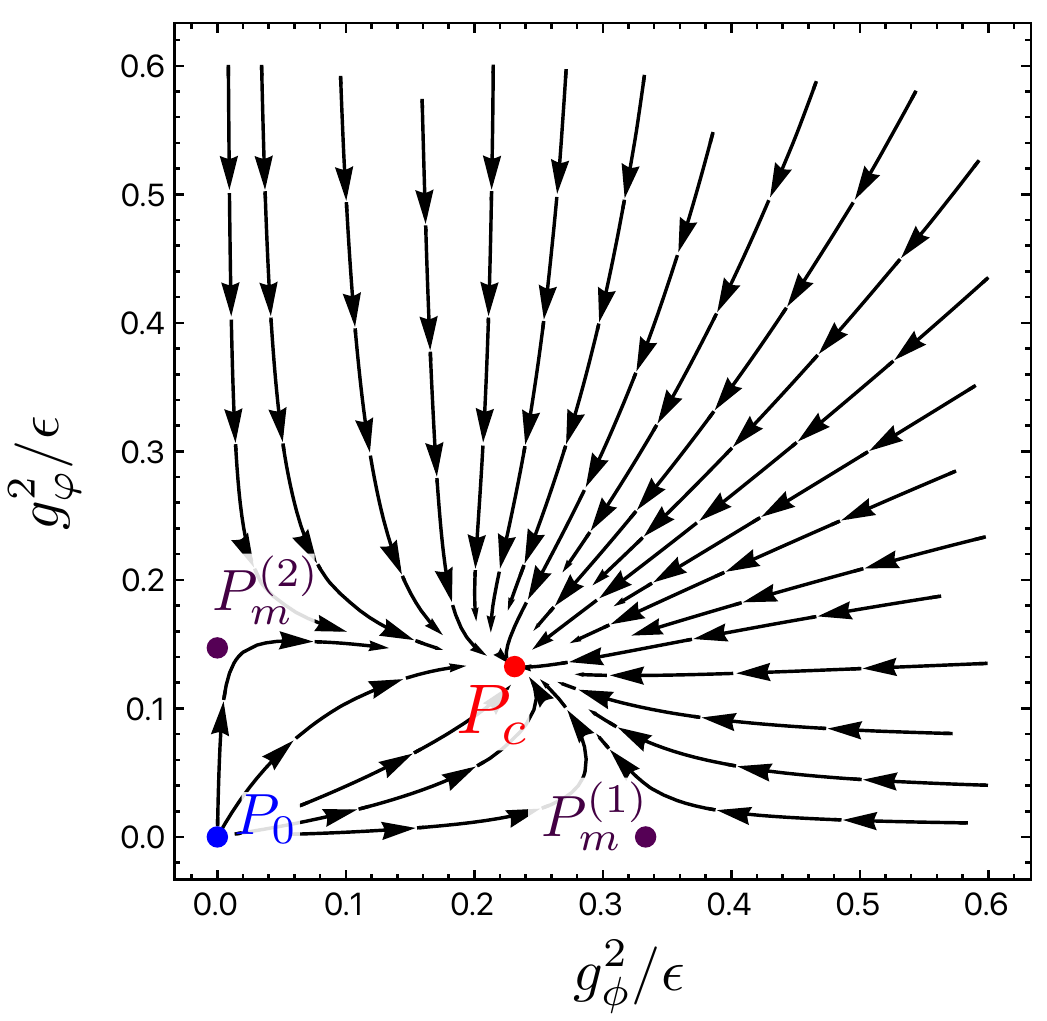}
    \caption{RG flow of the intra- and inter-valley Yukawa couplings $g_\phi$ and $g_\varphi$ for $N_\psi=3$ valleys. For other $N_\psi$ the same qualitative behavior is observed. The critical GNY 
    fixed point $P_m^{(1)}$ in the GNY theory of decoupled valleys is unstable against inter-valley coupling, leading to the formation of a new critical fixed point $P_c$.}  
    \label{figure4}
\end{figure}

For any given number $N_\psi\ge 3$ of Dirac fermion valleys with a corresponding angle $\theta = \pi/N_\psi$ between adjacent valleys, we numerically determine the fixed points of the 
four coupled RG equations. The values of $c_\phi^2$, $c_\varphi^2$, $g_\phi^2$ and $g_\varphi^2$ at the critical fixed point are then inserted into Eqs. (\ref{eq.z}), (\ref{eq.etapsi}), (\ref{eq.etaphi})
and (\ref{eq.etavarphi}) to determine the dynamical critical exponent $z$ and the anomalous dimensions $\eta_\psi$, $\eta_\phi$ and $\eta_\varphi$ of the fermion and boson fields. 

In addition to the non-interacting fixed point $P_0$ we find two metastable critical points, $P_m^{(1)} (g_\phi^2\neq 0, g_\varphi^2=0)$ and $P_m^{(2)}(g_\phi^2=0, g_\varphi^2\neq 0)$,
and a stable critical point $P_c(g_\phi^2\neq 0, g_\varphi^2\neq 0)$, as illustrated in Fig.~\ref{figure4} for $N_\psi=3$. The  behavior remains qualitatively the same for larger values of $N_\psi$.

\begin{figure}[t!]
    \centering
    \includegraphics[width=\linewidth]{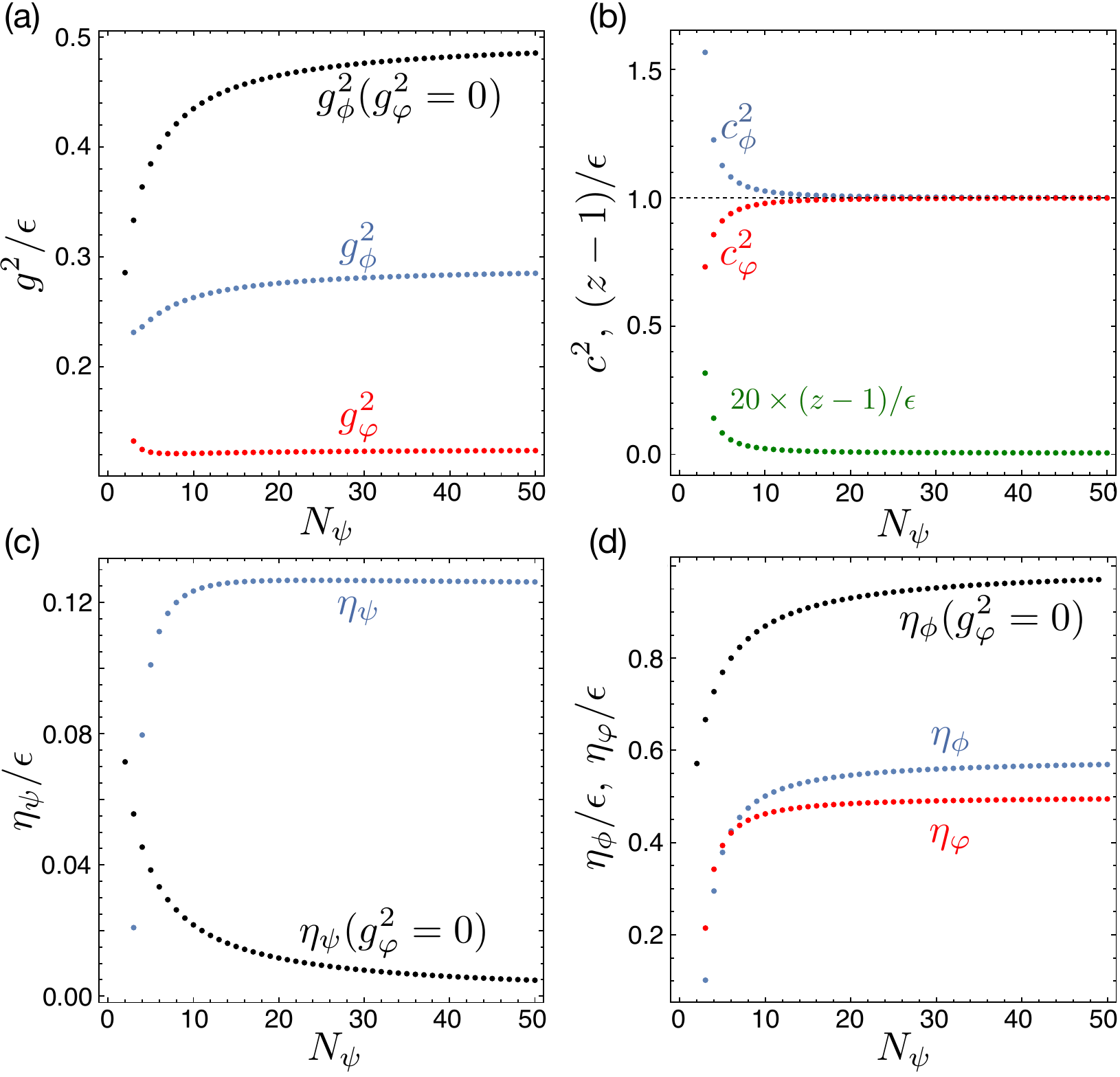}
    \caption{(a) Intra- and inter-valley Yukawa coupling constants $g_\phi^2$ and $g_\varphi^2$ at the critical fixed point $P_c$ as a function of the number $N_\psi$ of Dirac valleys. For comparison, the $N_\psi$ 
    dependence of the critical Yukawa $g_\phi^2 (g_\varphi^2=0)$ in a theory of decoupled valleys is shown in black. (b) The dynamical exponent $z$ and velocities $c_\phi$ and $c_\varphi$ of intra- and inter-valley 
    fluctuations as a function of $N_\psi$. Note that we have rescaled the order $\epsilon$ correction to $z$ by a factor of 20. The anomalous dimensions $\eta_\psi$ of the fermion field and $\eta_\phi$ and $\eta_\varphi$ of the bosonic fluctuations 
    fields are shown in panels (c) and (d), respectively.}
    \label{figure5}
\end{figure}

In Fig.~\ref{figure5}(a) the values of the Yukawa couplings $g_\phi^2$, $g_\varphi^2$ at the critical point $P_c$ are shown as a function of $N_\psi$, in comparison to the value $g_\phi^2$ at the GNY fixed point $P_m^{(1)}$ of 
 decoupled valleys ($g_\varphi^2=0$). For large $N_\psi$, we obtain analytic expressions up to order $1/N_\psi$, 
 \begin{equation}
  g_{\phi}^2 \simeq  {\frac{7\epsilon}{24}}\left(1-\frac{83}{72N_\psi} \right),\;\;  g_\varphi^2 \simeq  {\frac{\epsilon}{8}} \left(1-\frac{35}{72N_\psi} \right).
 \end{equation}

As shown in Fig.~\ref{figure5}(b), the velocities at $P_c$ take values $c_\phi^2>1$ and $c_\varphi^2<1$ and the dynamical exponent $z>1$ for all values of $N_\psi$, indicating a breaking of 
 Lorentz invariance. In the limit $N_\psi\to\infty$, Lorentz invariance is restored, $c_\phi^2,c_\varphi^2\to 1$ and $z\to 1$. Note that the corrections to the asymptotic values are of order $1/N_\psi^2$ and 
 arise from the expansion of the factors $\cos(\theta)=\cos(\pi/N_\psi)$.  This shows that the breaking of Lorentz invariance is a consequence of the misalignment of the coordinate frames of 
 neighboring Dirac valleys.
 
 The $N_\psi$ dependence of the anomalous dimension $\eta_\psi$ of the fermion field and of the anomalous dimensions $\eta_\phi$ and $\eta_\varphi$ of the fluctuation fields at $P_c$ is shown in Figs.~\ref{figure5}
 (c) and (d), respectively. In the limit $N_\psi \to \infty$ the anomalous dimensions approach finite, non-zero values, given by the expressions
 \begin{equation}
  \eta_\psi \simeq  {\frac{\epsilon}{8}}\left(1-\frac{49}{72N_\psi} \right)
 \end{equation}
for the fermion anomalous dimension and 
 \begin{equation}
\eta_\phi \simeq   {\frac{7\epsilon}{12}}\left(1-\frac{83}{72N_\psi} \right), \;
\eta_\varphi \simeq {\frac{\epsilon}{2}} \left(1-\frac{35}{72N_\psi} \right)
 \end{equation}
for the boson anomalous dimensions.

\begin{figure}[t!]
    \centering
    \includegraphics[width=\linewidth]{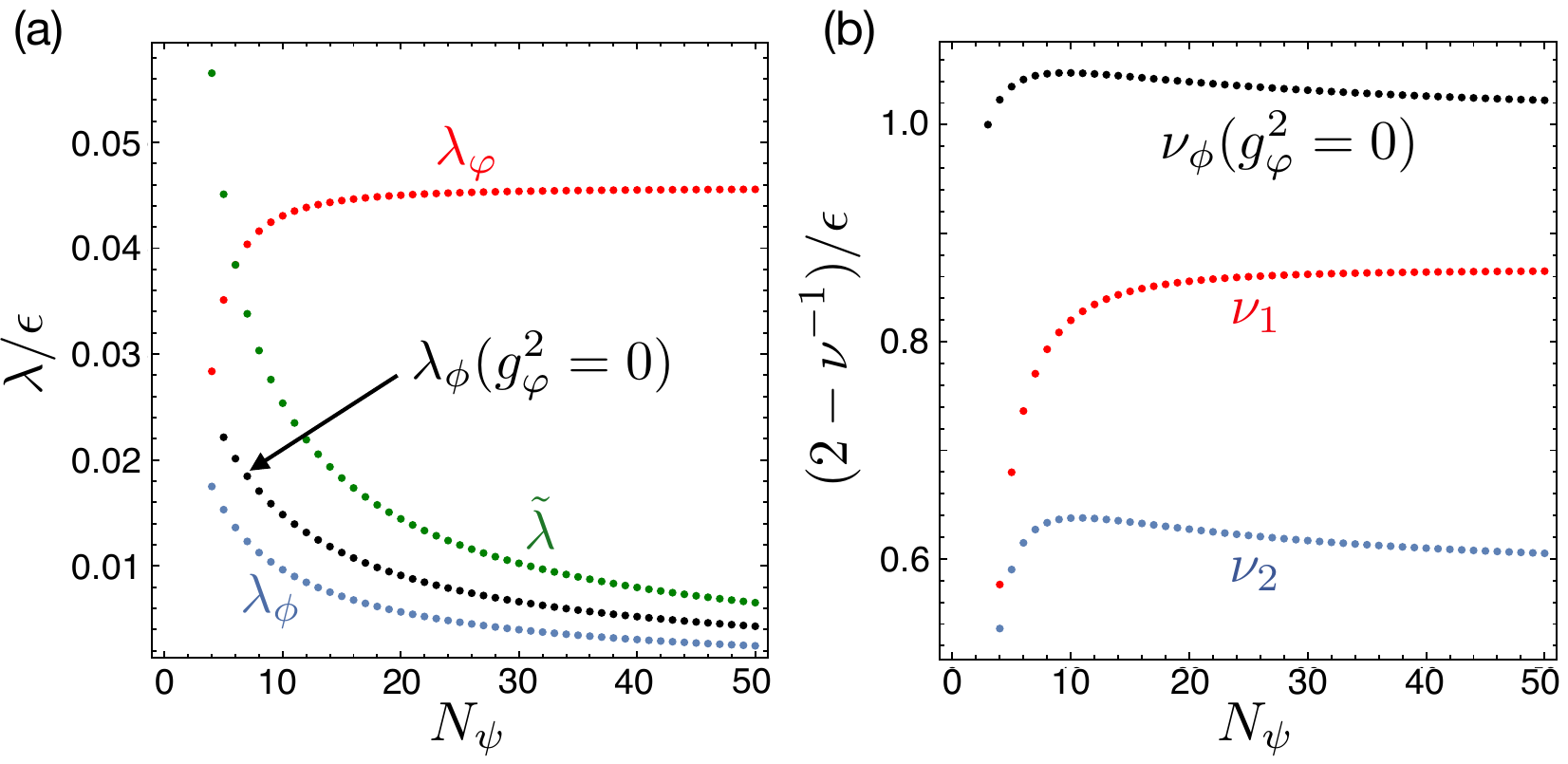}
    \caption{(a) Critical values of the bosonic interactions $\lambda_\phi$, $\lambda_\varphi$, and $\tilde{\lambda}$ at the GNY fixed point $P_c$ as a function of $N_\psi$. For comparison, the 
    coupling strength $\lambda_\phi(g_\varphi^2=0)$ of the $\phi^4$ vertex in the GNY theory with decoupled valleys is shown in black. (b) The correlation length exponents $\nu_1$ and $\nu_2$ as a function of 
    $N_\psi$.}
    \label{figure6}
\end{figure}

We now turn our attention to the coupled RG equations (\ref{eq.lambda1}), (\ref{eq.lambda2}) and (\ref{eq.lambda3}) for the bosonic vertices $\lambda_\phi$, $\lambda_\varphi$ and $\tilde{\lambda}$.
For any given $N_\psi$, we insert the numerically determined values of the Yukawa couplings $g_\phi^2$, $g_\varphi^2$ and boson velocities $c_\phi^2$, $c_\varphi^2$ at the stable critical point $P_c$, as well as the resulting value of the dynamical 
exponent $z$. We then numerically determine all non-trivial fixed points for $(\lambda_\phi, \lambda_\varphi,\tilde{\lambda})$ and analyze their stability. For $N_\psi \ge 4$ we find one stable fixed point of the boson interactions. However, for 
$N_\psi=3$ the stable fixed point is lost, which could be an artifact of the perturbative RG treatment and may change at higher-loop order. The runaway flow could also indicate first-order behavior. 

In Fig.~\ref{figure6}(a), the values of $\lambda_\phi$, $\lambda_\varphi$ and $\tilde{\lambda}$ are shown as a function of $N_\psi\ge 4$ and compared to the value of $\lambda_\phi$ in the GNY theory of decoupled 
Dirac valleys ($g_\varphi^2=0$). Their asymptotic behavior for large $N_\psi$ is described by the expressions 
\begin{eqnarray}
 \lambda_\phi &  \simeq & \frac{245(4\sqrt{30}-11)\epsilon}{96(5+2\sqrt{30})^2 N_\psi},\\
  \lambda_\varphi & \simeq & \frac{\epsilon}{4\sqrt{30}}\left[ 1-\frac{7(30-6\sqrt{30})}{432 N_\psi}\right],\\
     \tilde\lambda & \simeq & \frac{35 \epsilon}{6(5+2\sqrt{30})N_\psi}.
\end{eqnarray}

Finally, for $N_\psi\ge 4$, where a stable critical point exists, we compute the inverse correlation length exponents $\nu_1^{-1}$ and $\nu_2^{-1}$ of  bosonic fluctuations
from the eigenvalues of the 2x2 matrix in Eq.~(\ref{eq.mass}), evaluated at the critical point. The resulting behavior is shown in Fig.~\ref{figure6}(b) and the asymptotic forms are given by
\begin{eqnarray}
    \nu^{-1}_{1} & = & 2-\epsilon \left[\frac{15+2\sqrt{30}}{30}-\frac{7(3+2\sqrt{30})}{432 N_\psi}\right],\\
    \nu^{-1}_{2} & = & 2-\epsilon \left[ \frac{7}{12}-\frac{7(634007+120960)}{311904 N_\psi}\right].
\end{eqnarray}

\section{Discussion}
\label{sec.discussion}

In this work, we have investigated the effects of inter-valley coupling on the quantum criticality of Dirac fermions. Interactions between different Dirac valleys in momentum space are unavoidable since phase transitions 
are typically driven by short range interactions in an underlying lattice model \cite{Grushin+13}. Our work was motivated by the occurrence of a field-driven topological phase transition in the Kitaev QSL, at which Dirac points of 
emergent fermions form at the three $M$ points in the Brillouin zone  \cite{Ralko+20,Yilmaz+22,Zhang+2022,Thiagarajan+2026}. 

We have analyzed an Ising GNY theory with intra- and inter-valley fluctuations, using a momentum shell RG and performing an $\epsilon$ expansion below four space-time dimensions to one-loop order. In contrast to the
previous large-$N$ RG approach in 2+1 dimensions \cite{Thiagarajan+2026}, where a GNY fixed point with finite inter-valley coupling is postulated and scale invariance of the Yukawa couplings at criticality utilized,  the present $4-\epsilon$ expansion allows 
for an unbiased analysis of the scale dependence of intra- and inter-valley Yukawa couplings. 

 We have generalized to a situation of $N_\psi\ge 3$ Dirac valleys arranged on a ring in momentum space, with a relative angle $\theta=\pi/N_\psi$ between the coordinate frames of neighboring valleys. For simplicity, 
 we have restricted inter-valley interactions to adjacent valleys to reduce the number of coupling constants. We have numerically determined fixed points and critical exponents for any given value of $N_\psi$ and derived
 analytic expressions in the large-$N_\psi$ limit up to order $1/N_\psi$.
 
 In the absence of inter-valley coupling, $g_\varphi = 0$, we recover the standard $4-\epsilon$ results for the Ising GNY theory \cite{Herbut+2006,Herbut+2009,Justin+1991}. However, our results show that the critical fixed point 
 in the GNY theory of decoupled valleys is unstable against small inter-valley coupling, resulting in a new critical fixed point $P_c$ with finite intra- and inter-valley Yukawa couplings. 
 
 At $P_c$ the critical exponents are  significantly different from those of the conventional GNY theory. Most importantly, the dynamical exponent acquires a value $z>1$, indicating the breaking of Lorentz invariance. This is a result of interference 
 effects from the relative rotation of coordinate frames of adjacent valleys. In the large-$N_\psi$ limit, Lorentz-violating effects are progressively suppressed, and the velocities of the bosonic fluctuation fields as well as the dynamical exponent
 asymptotically approach 1.  In this limit, we further find that both Yukawa couplings flow to finite values, while only the inter-valley quartic vertex remains finite, with all other quartic couplings vanishing. The critical exponents 
approach finite values that differ from those of the purely intra-valley GNY theory \cite{Justin+1991,Herbut+2006,Herbut+2009}. This demonstrates that the inclusion of inter-valley couplings leads to a 
 universality class distinct from that of the standard Ising GNY theory.

In future studies it would be interesting to investigate the effects of inter-valley coupling at higher-loop order and to consider different order-parameter symmetries. In the context of U(1) Dirac QSL one could also include the gauge fields 
via minimal coupling and study their interplay with intra- and inter-valley fluctuations.


\begin{thebibliography}{19}%
\makeatletter
\providecommand \@ifxundefined [1]{%
 \@ifx{#1\undefined}
}%
\providecommand \@ifnum [1]{%
 \ifnum #1\expandafter \@firstoftwo
 \else \expandafter \@secondoftwo
 \fi
}%
\providecommand \@ifx [1]{%
 \ifx #1\expandafter \@firstoftwo
 \else \expandafter \@secondoftwo
 \fi
}%
\providecommand \natexlab [1]{#1}%
\providecommand \enquote  [1]{``#1''}%
\providecommand \bibnamefont  [1]{#1}%
\providecommand \bibfnamefont [1]{#1}%
\providecommand \citenamefont [1]{#1}%
\providecommand \href@noop [0]{\@secondoftwo}%
\providecommand \href [0]{\begingroup \@sanitize@url \@href}%
\providecommand \@href[1]{\@@startlink{#1}\@@href}%
\providecommand \@@href[1]{\endgroup#1\@@endlink}%
\providecommand \@sanitize@url [0]{\catcode `\\12\catcode `\$12\catcode
  `\&12\catcode `\#12\catcode `\^12\catcode `\_12\catcode `\%12\relax}%
\providecommand \@@startlink[1]{}%
\providecommand \@@endlink[0]{}%
\providecommand \url  [0]{\begingroup\@sanitize@url \@url }%
\providecommand \@url [1]{\endgroup\@href {#1}{\urlprefix }}%
\providecommand \urlprefix  [0]{URL }%
\providecommand \Eprint [0]{\href }%
\providecommand \doibase [0]{https://doi.org/}%
\providecommand \selectlanguage [0]{\@gobble}%
\providecommand \bibinfo  [0]{\@secondoftwo}%
\providecommand \bibfield  [0]{\@secondoftwo}%
\providecommand \translation [1]{[#1]}%
\providecommand \BibitemOpen [0]{}%
\providecommand \bibitemStop [0]{}%
\providecommand \bibitemNoStop [0]{.\EOS\space}%
\providecommand \EOS [0]{\spacefactor3000\relax}%
\providecommand \BibitemShut  [1]{\csname bibitem#1\endcsname}%
\let\auto@bib@innerbib\@empty
\bibitem [{\citenamefont {El-Batanouny}(2020)}]{El-Batanouny+2020}%
  \BibitemOpen
  \bibfield  {author} {\bibinfo {author} {\bibfnamefont {M.}~\bibnamefont
  {El-Batanouny}},\ }\bibinfo {title} {Dirac materials and dirac fermions},\
  in\ \href@noop {} {\emph {\bibinfo {booktitle} {Advanced Quantum Condensed
  Matter Physics: One-Body, Many-Body, and Topological Perspectives}}}\
  (\bibinfo  {publisher} {Cambridge University Press},\ \bibinfo {year}
  {2020})\ pp.\ \bibinfo {pages} {331--372}\BibitemShut {NoStop}%
\bibitem [{\citenamefont {Novoselov}\ \emph {et~al.}(2005)\citenamefont
  {Novoselov}, \citenamefont {Geim}, \citenamefont {Morozov}, \citenamefont
  {Jiang}, \citenamefont {Katsnelson}, \citenamefont {Grigorieva},
  \citenamefont {Dubonos},\ and\ \citenamefont {Firsov}}]{Novoselov+2005}%
  \BibitemOpen
  \bibfield  {author} {\bibinfo {author} {\bibfnamefont {K.~S.}\ \bibnamefont
  {Novoselov}}, \bibinfo {author} {\bibfnamefont {A.~K.}\ \bibnamefont {Geim}},
  \bibinfo {author} {\bibfnamefont {S.~V.}\ \bibnamefont {Morozov}}, \bibinfo
  {author} {\bibfnamefont {D.}~\bibnamefont {Jiang}}, \bibinfo {author}
  {\bibfnamefont {M.~I.}\ \bibnamefont {Katsnelson}}, \bibinfo {author}
  {\bibfnamefont {I.~V.}\ \bibnamefont {Grigorieva}}, \bibinfo {author}
  {\bibfnamefont {S.~V.}\ \bibnamefont {Dubonos}},\ and\ \bibinfo {author}
  {\bibfnamefont {A.~A.}\ \bibnamefont {Firsov}},\ }\href
  {https://doi.org/10.1038/nature04233} {\bibfield  {journal} {\bibinfo
  {journal} {Nature}\ }\textbf {\bibinfo {volume} {438}},\ \bibinfo {pages}
  {197} (\bibinfo {year} {2005})}\BibitemShut {NoStop}%
\bibitem [{\citenamefont {Armitage}\ \emph {et~al.}(2018)\citenamefont
  {Armitage}, \citenamefont {Mele},\ and\ \citenamefont
  {Vishwanath}}]{Armitage+18}%
  \BibitemOpen
  \bibfield  {author} {\bibinfo {author} {\bibfnamefont {N.~P.}\ \bibnamefont
  {Armitage}}, \bibinfo {author} {\bibfnamefont {E.~J.}\ \bibnamefont {Mele}},\
  and\ \bibinfo {author} {\bibfnamefont {A.}~\bibnamefont {Vishwanath}},\
  }\href {https://doi.org/10.1103/RevModPhys.90.015001} {\bibfield  {journal}
  {\bibinfo  {journal} {Rev. Mod. Phys.}\ }\textbf {\bibinfo {volume} {90}},\
  \bibinfo {pages} {015001} (\bibinfo {year} {2018})}\BibitemShut {NoStop}%
\bibitem [{\citenamefont {Wen}(2002)}]{Wen02}%
  \BibitemOpen
  \bibfield  {author} {\bibinfo {author} {\bibfnamefont {X.-G.}\ \bibnamefont
  {Wen}},\ }\href {https://doi.org/10.1103/PhysRevB.65.165113} {\bibfield
  {journal} {\bibinfo  {journal} {Phys. Rev. B}\ }\textbf {\bibinfo {volume}
  {65}},\ \bibinfo {pages} {165113} (\bibinfo {year} {2002})}\BibitemShut
  {NoStop}%
\bibitem [{\citenamefont {Kitaev}(2006)}]{Kitaev06}%
  \BibitemOpen
  \bibfield  {author} {\bibinfo {author} {\bibfnamefont {A.}~\bibnamefont
  {Kitaev}},\ }\href {https://doi.org/10.1016/j.aop.2005.10.005} {\bibfield
  {journal} {\bibinfo  {journal} {Annals of Physics}\ }\textbf {\bibinfo
  {volume} {321}},\ \bibinfo {pages} {2–111} (\bibinfo {year}
  {2006})}\BibitemShut {NoStop}%
\bibitem [{\citenamefont {Grushin}\ \emph {et~al.}(2013)\citenamefont
  {Grushin}, \citenamefont {Castro}, \citenamefont {Cortijo}, \citenamefont
  {de~Juan}, \citenamefont {Vozmediano},\ and\ \citenamefont
  {Valenzuela}}]{Grushin+13}%
  \BibitemOpen
  \bibfield  {author} {\bibinfo {author} {\bibfnamefont {A.~G.}\ \bibnamefont
  {Grushin}}, \bibinfo {author} {\bibfnamefont {E.~V.}\ \bibnamefont {Castro}},
  \bibinfo {author} {\bibfnamefont {A.}~\bibnamefont {Cortijo}}, \bibinfo
  {author} {\bibfnamefont {F.}~\bibnamefont {de~Juan}}, \bibinfo {author}
  {\bibfnamefont {M.~A.~H.}\ \bibnamefont {Vozmediano}},\ and\ \bibinfo
  {author} {\bibfnamefont {B.}~\bibnamefont {Valenzuela}},\ }\href
  {https://doi.org/10.1103/PhysRevB.87.085136} {\bibfield  {journal} {\bibinfo
  {journal} {Phys. Rev. B}\ }\textbf {\bibinfo {volume} {87}},\ \bibinfo
  {pages} {085136} (\bibinfo {year} {2013})}\BibitemShut {NoStop}%
\bibitem [{\citenamefont {Gross}\ and\ \citenamefont
  {Neveu}(1974)}]{Gross+1974}%
  \BibitemOpen
  \bibfield  {author} {\bibinfo {author} {\bibfnamefont {D.~J.}\ \bibnamefont
  {Gross}}\ and\ \bibinfo {author} {\bibfnamefont {A.}~\bibnamefont {Neveu}},\
  }\href {https://doi.org/10.1103/PhysRevD.10.3235} {\bibfield  {journal}
  {\bibinfo  {journal} {Phys. Rev. D}\ }\textbf {\bibinfo {volume} {10}},\
  \bibinfo {pages} {3235} (\bibinfo {year} {1974})}\BibitemShut {NoStop}%
\bibitem [{\citenamefont {Zinn-Justin}(1991)}]{Justin+1991}%
  \BibitemOpen
  \bibfield  {author} {\bibinfo {author} {\bibfnamefont {J.}~\bibnamefont
  {Zinn-Justin}},\ }\href
  {https://doi.org/https://doi.org/10.1016/0550-3213(91)90043-W} {\bibfield
  {journal} {\bibinfo  {journal} {Nuclear Physics B}\ }\textbf {\bibinfo
  {volume} {367}},\ \bibinfo {pages} {105} (\bibinfo {year}
  {1991})}\BibitemShut {NoStop}%
\bibitem [{\citenamefont {Herbut}(2006)}]{Herbut+2006}%
  \BibitemOpen
  \bibfield  {author} {\bibinfo {author} {\bibfnamefont {I.~F.}\ \bibnamefont
  {Herbut}},\ }\href {https://doi.org/10.1103/PhysRevLett.97.146401} {\bibfield
   {journal} {\bibinfo  {journal} {Phys. Rev. Lett.}\ }\textbf {\bibinfo
  {volume} {97}},\ \bibinfo {pages} {146401} (\bibinfo {year}
  {2006})}\BibitemShut {NoStop}%
\bibitem [{\citenamefont {Herbut}\ \emph {et~al.}(2009)\citenamefont {Herbut},
  \citenamefont {Juri},\ and\ \citenamefont {Vafek}}]{Herbut+2009}%
  \BibitemOpen
  \bibfield  {author} {\bibinfo {author} {\bibfnamefont {I.~F.}\ \bibnamefont
  {Herbut}}, \bibinfo {author} {\bibfnamefont {V.}~\bibnamefont {Juri}},\ and\
  \bibinfo {author} {\bibfnamefont {O.}~\bibnamefont {Vafek}},\ }\href
  {https://doi.org/10.1103/PhysRevB.80.075432} {\bibfield  {journal} {\bibinfo
  {journal} {Phys. Rev. B}\ }\textbf {\bibinfo {volume} {80}},\ \bibinfo
  {pages} {075432} (\bibinfo {year} {2009})}\BibitemShut {NoStop}%
\bibitem [{\citenamefont {Janssen}\ and\ \citenamefont
  {Herbut}(2014)}]{Herbut+2014}%
  \BibitemOpen
  \bibfield  {author} {\bibinfo {author} {\bibfnamefont {L.}~\bibnamefont
  {Janssen}}\ and\ \bibinfo {author} {\bibfnamefont {I.~F.}\ \bibnamefont
  {Herbut}},\ }\href {https://doi.org/10.1103/PhysRevB.89.205403} {\bibfield
  {journal} {\bibinfo  {journal} {Phys. Rev. B}\ }\textbf {\bibinfo {volume}
  {89}},\ \bibinfo {pages} {205403} (\bibinfo {year} {2014})}\BibitemShut
  {NoStop}%
\bibitem [{\citenamefont {Nielsen}\ and\ \citenamefont
  {Ninomiya}(1981)}]{Nielsen+1981}%
  \BibitemOpen
  \bibfield  {author} {\bibinfo {author} {\bibfnamefont {H.~B.}\ \bibnamefont
  {Nielsen}}\ and\ \bibinfo {author} {\bibfnamefont {M.}~\bibnamefont
  {Ninomiya}},\ }\href
  {https://doi.org/https://doi.org/10.1016/0370-2693(81)91026-1} {\bibfield
  {journal} {\bibinfo  {journal} {Physics Letters B}\ }\textbf {\bibinfo
  {volume} {105}},\ \bibinfo {pages} {219} (\bibinfo {year}
  {1981})}\BibitemShut {NoStop}%
\bibitem [{\citenamefont {Suzuki}(2004)}]{Suzuki+2004}%
  \BibitemOpen
  \bibfield  {author} {\bibinfo {author} {\bibfnamefont {H.}~\bibnamefont
  {Suzuki}},\ }\href {https://doi.org/10.1143/PTP.112.855} {\bibfield
  {journal} {\bibinfo  {journal} {Progress of Theoretical Physics}\ }\textbf
  {\bibinfo {volume} {112}},\ \bibinfo {pages} {855} (\bibinfo {year}
  {2004})}\BibitemShut {NoStop}%
\bibitem [{\citenamefont {Drell}\ \emph {et~al.}(1976)\citenamefont {Drell},
  \citenamefont {Weinstein},\ and\ \citenamefont {Yankielowicz}}]{Drell+1976}%
  \BibitemOpen
  \bibfield  {author} {\bibinfo {author} {\bibfnamefont {S.~D.}\ \bibnamefont
  {Drell}}, \bibinfo {author} {\bibfnamefont {M.}~\bibnamefont {Weinstein}},\
  and\ \bibinfo {author} {\bibfnamefont {S.}~\bibnamefont {Yankielowicz}},\
  }\href {https://doi.org/10.1103/PhysRevD.14.1627} {\bibfield  {journal}
  {\bibinfo  {journal} {Phys. Rev. D}\ }\textbf {\bibinfo {volume} {14}},\
  \bibinfo {pages} {1627} (\bibinfo {year} {1976})}\BibitemShut {NoStop}%
\bibitem [{\citenamefont {Lang}\ and\ \citenamefont
  {L\"auchli}(2019)}]{Lang+19}%
  \BibitemOpen
  \bibfield  {author} {\bibinfo {author} {\bibfnamefont {T.~C.}\ \bibnamefont
  {Lang}}\ and\ \bibinfo {author} {\bibfnamefont {A.~M.}\ \bibnamefont
  {L\"auchli}},\ }\href {https://doi.org/10.1103/PhysRevLett.123.137602}
  {\bibfield  {journal} {\bibinfo  {journal} {Phys. Rev. Lett.}\ }\textbf
  {\bibinfo {volume} {123}},\ \bibinfo {pages} {137602} (\bibinfo {year}
  {2019})}\BibitemShut {NoStop}%
\bibitem [{\citenamefont {Zhang}\ \emph {et~al.}(2022)\citenamefont {Zhang},
  \citenamefont {Hal\'asz},\ and\ \citenamefont {Batista}}]{Zhang+2022}%
  \BibitemOpen
  \bibfield  {author} {\bibinfo {author} {\bibfnamefont {S.-S.}\ \bibnamefont
  {Zhang}}, \bibinfo {author} {\bibfnamefont {G.~B.}\ \bibnamefont
  {Hal\'asz}},\ and\ \bibinfo {author} {\bibfnamefont {C.~D.}\ \bibnamefont
  {Batista}},\ }\href@noop {} {\bibfield  {journal} {\bibinfo  {journal}
  {Nature Communications}\ }\textbf {\bibinfo {volume} {13}},\ \bibinfo {pages}
  {399} (\bibinfo {year} {2022})}\BibitemShut {NoStop}%
\bibitem [{\citenamefont {Thiagarajan}\ \emph {et~al.}(2026)\citenamefont
  {Thiagarajan}, \citenamefont {Watson}, \citenamefont {Yzeiri}, \citenamefont
  {Hu}, \citenamefont {Uchoa},\ and\ \citenamefont
  {Kr\"uger}}]{Thiagarajan+2026}%
  \BibitemOpen
  \bibfield  {author} {\bibinfo {author} {\bibfnamefont {S.}~\bibnamefont
  {Thiagarajan}}, \bibinfo {author} {\bibfnamefont {C.}~\bibnamefont {Watson}},
  \bibinfo {author} {\bibfnamefont {T.}~\bibnamefont {Yzeiri}}, \bibinfo
  {author} {\bibfnamefont {H.}~\bibnamefont {Hu}}, \bibinfo {author}
  {\bibfnamefont {B.}~\bibnamefont {Uchoa}},\ and\ \bibinfo {author}
  {\bibfnamefont {F.}~\bibnamefont {Kr\"uger}},\ }\href
  {https://doi.org/10.1103/jjj6-cx8l} {\bibfield  {journal} {\bibinfo
  {journal} {Phys. Rev. B}\ }\textbf {\bibinfo {volume} {113}},\ \bibinfo
  {pages} {085108} (\bibinfo {year} {2026})}\BibitemShut {NoStop}%
\bibitem [{\citenamefont {Ralko}\ and\ \citenamefont
  {Merino}(2020)}]{Ralko+20}%
  \BibitemOpen
  \bibfield  {author} {\bibinfo {author} {\bibfnamefont {A.}~\bibnamefont
  {Ralko}}\ and\ \bibinfo {author} {\bibfnamefont {J.}~\bibnamefont {Merino}},\
  }\href {https://doi.org/10.1103/PhysRevLett.124.217203} {\bibfield  {journal}
  {\bibinfo  {journal} {Phys. Rev. Lett.}\ }\textbf {\bibinfo {volume} {124}},\
  \bibinfo {pages} {217203} (\bibinfo {year} {2020})}\BibitemShut {NoStop}%
\bibitem [{\citenamefont {{Y\ifmmode \imath \else \i \fi{}lmaz, F. and Kampf,
  A. P. and Yip, S. K.}}(2022)}]{Yilmaz+22}%
  \BibitemOpen
  \bibfield  {author} {\bibinfo {author} {\bibnamefont {{Y\ifmmode \imath \else
  \i \fi{}lmaz, F. and Kampf, A. P. and Yip, S. K.}}},\ }\href
  {https://doi.org/10.1103/PhysRevResearch.4.043024} {\bibfield  {journal}
  {\bibinfo  {journal} {Phys. Rev. Res.}\ }\textbf {\bibinfo {volume} {4}},\
  \bibinfo {pages} {043024} (\bibinfo {year} {2022})}\BibitemShut {NoStop}%
\end{thebibliography}
\end{document}